# High Quality Factor Fano-Resonant All-Dielectric Metamaterials

Yuanmu Yang[1], Ivan I. Kravchenko[2], Dayrl P. Briggs[2], Jason Valentine[3*]

[1]*Interdisciplinary Materials Science Program, Vanderbilt University, Nashville, Tennessee 37212, USA*
[2]*Center for Nanophase Materials Sciences, Oak Ridge National Laboratory, Oak Ridge, Tennessee 37831, USA*
[3]*Department of Mechanical Engineering, Vanderbilt University, Nashville, Tennessee 37212, USA*
[*]email: jason.g.valentine@vanderbilt.edu

**Metasurface analogues of electromagnetically induced transparency (EIT) have been a major focus of the nanophotonics field over the past several years due their ability to produce high quality factor (Q-factor) resonances for applications such as low-loss slow light devices and highly sensitive optical sensors. However, Ohmic loss limits the achievable Q-factors in conventional plasmonic EIT metasurfaces to values less than ~10, significantly hampering device performance. Here, we report experimental demonstration of a classical analogue of EIT using all-dielectric silicon-based metasurfaces. Due to extremely low absorption loss and coherent interaction of neighboring "meta-atoms", a record-high Q-factor of 483 is experimentally observed, leading to a refractive index sensor with a figure-of-merit (FOM) of 103. Furthermore, we show that the dielectric metasurfaces can be engineered to confine the optical field in either the silicon resonator or the environment, allowing one to tailor light-matter interaction at the nanoscale.**

Electromagnetically induced transparency (EIT) is a concept originally observed in atomic physics and arises due to quantum interference resulting in a narrowband transparency window for light propagating through an originally opaque medium[1]. This concept was later extended to classical optical systems using plasmonic metamaterials[2–10], among others[11,12], allowing experimental implementation with incoherent light and operation at room temperature. The transparent and highly dispersive nature of EIT offers a potential solution to the long-standing issue of loss in metamaterials as well as the creation of ultra-high quality factor (Q-factor) resonances, which are critical for realizing low-loss slow light devices[2,3,6,10], optical sensors[13,14] and enhancing nonlinear interactions[15].

The classical analogue of EIT in plasmonic metamaterials relies on a Fano-type interference[16,17] between a broadband "bright" mode resonator, that is accessible from free space, and a narrowband "dark" mode resonator which is less-accessible, or inaccessible, from free-space. If these two resonances are brought in close proximity in both the spatial and frequency domains, they can interfere resulting in an extremely narrow reflection or transmission window. Due to the low radiative loss of the dark mode, the Fano resonance can be extremely sharp, resulting in complete transmission, analogous to EIT[2–10], or complete reflection[13], from the sample across a very narrow bandwidth. However, the main limitation of metal-based Fano-resonant systems is the large non-radiative loss due to Drude damping, which limits the achievable Q-factor[17] to less than ~10.

High-refractive-index dielectric particles offer a potential solution to the issue of material (non-radiative) loss. Such particles exhibit magnetic and electric dipole, and higher order, Mie resonances while suffering from minimal absorption loss, provided the illumination energy is sub-bandgap[18–22]. For instance, dielectric Fano-resonant structures based on oligomer antennae[23], silicon nanostripe[24], and asymmetric cut-wire metamaterials[25,26] have been demonstrated with Q-factors up to 127.

In this article, we describe the development of silicon (Si)-based metasurfaces possessing sharp EIT-like resonances with a record high Q-factor of 483 in the near-infrared regime. The high-Q resonance is accomplished by employing Fano-resonant unit cells in which both radiative and non-radiative damping are minimized through coherent interaction among the resonators combined with the reduction of absorption loss. Combining the narrow resonance linewidth with strong near field confinement, we demonstrate an optical refractive index sensor with a figure of merit (*FOM)* of 103. In addition, we demonstrate unit cell designs consisting of double-gap split-ring resonators that possess narrow feed-gaps in which the electric field can be further enhanced in the surrounding medium, allowing interaction with emitters such as quantum dots.

**Results**

**Design and Characterization.** The schematic of the designed dielectric metasurface is shown in Fig. 1a. The structure is formed from a periodic lattice made of a rectangular bar resonator and a ring resonator, both formed from Si. The rectangular bar resonator serves as an electric dipole antenna which couples strongly to free space excitation with the incident E-field oriented along the *x*-axis. The collective oscillations of the bar resonators form the "bright" mode resonance. The ring supports a magnetic dipole mode wherein the electric field is directed along the azimuthal direction, rotating around the ring's axis. The magnetic dipole mode in the ring cannot be directly excited by light at normal incidence as the magnetic arm of the incident wave is perpendicular to the dipole axis; however, it can couple to the bright mode bar resonator. Furthermore, the ring resonators interact through near-field coupling, resulting in collective oscillation of the resonators and suppression of radiative loss, forming the "dark" mode of the system. The interference between the collective bright and dark modes form a typical 3-level Fano-resonant system[17], as illustrated in Fig. 1b. The response of the dielectric metasurface is similar to the collective modes found in asymmetric double-gap split-ring resonators in which the asymmetry in the rings yields a finite electric dipole moment that can couple the out-of-plane magnetic dipole mode to free space[27–32]. In our case, coupling to free space is provided by the bright mode resonators which are placed in close proximity to the symmetric dark mode resonators. Numerical simulations of the structure were carried out using a commercially-available software (CST Microwave Studio) using the finite-element frequency-domain (FEFD) solver (see Supplementary Information section 1 for details of the simulation). The simulated transmittance, reflectance and absorption spectra of the designed structure are shown in Fig. 1c, where a distinct EIT-like peak can be observed at a wavelength of 1376 nm. In these simulations the resonators are sitting on a quartz substrate and embedded in a medium with a refractive index of $n = 1.44$, matching the experiments described below. We have also used the dielectric function of Si as determined using ellipsometry and assumed an infinitely large array, resulting in a Q-factor that reaches 1176, roughly 2 orders of magnitude higher than any previously reported Fano-resonant plasmonic metamaterials[17]. Furthermore, the peak of the transparency window approaches unity, demonstrating the potential to realize highly dispersive, yet lossless, "slow light" devices.

The designed structure was fabricated by starting with a 110-nm-thick poly-crystalline Si (refractive index $n=3.7$) thin film deposited on a quartz ($SiO_2$) ($n=1.48$) wafer. The structure was patterned using electron beam lithography for mask formation followed by reactive-ion etching. A scanning electron microscope (SEM) image of a fabricated sample is shown in Fig.

1d,e. Before optical measurements, the resonator array was immersed in a refractive index matching oil ($n=1.44$) within a polydimethylsiloxane (PDMS) flow cell. (See Methods for details). The experimentally measured transmittance, reflectance and absorption spectra, plotted in Fig. 1e, were acquired by illuminating the sample with normal-incident white light with the electric field oriented along the long axis of the bar resonator. A peak transmittance of 82% was observed at a wavelength of 1371 nm with a Q-factor of 483, as determined by fitting the dark mode resonance to a Fano line shape (see Supplementary Information section 2 for details). The shape of the measured spectra have good agreement with the simulation, though the Q-factor and peak transmittance are reduced. This is most likely due to imperfections within the fabricated sample which introduce scattering loss and break coherence among the resonators. The role of coherence will be further addressed below. Furthermore, additional loss in the Si arising from surface states created during reactive ion etching leads to slightly increased absorption in the array compared to theory.. Nonetheless, this structure has the highest experimentally achieved Q-factor among all plasmonic or dielectric optical metamaterials/metasurfaces reported to date.

**Theoretical Treatment.** The response of the metasurface can be qualitatively understand by applying the widely used coupled harmonic oscillator model[33,34], described by the following equations,

$$\dot{x}_1 - j(\omega_0 + j\gamma_1)x_1 + j\kappa x_2 = gE_0 e^{j\omega t},$$
$$\dot{x}_2 - j(\omega_0 + \delta + j\gamma_2)x_2 + j\kappa x_1 = 0,$$
(1)

where $x_1$ and $x_2$ represent the amplitude of the collective modes supported by oscillators 1 (bright mode) and 2 (dark mode), respectively. $\gamma_1$ and $\gamma_2$ are the damping rates, given by $\gamma = \gamma_R + \gamma_{NR}$ where $\gamma_R$ and $\gamma_{NR}$ are the radiative and non-radiative decay rates, respectively. $\omega_0$ is the central resonant frequency of oscillator 1, $\delta$ is the detuning of resonance frequency of oscillator 1 and 2, and $g$ is the bright mode dipole coupling strength to the incident electric field $E_0$. $\kappa$ is the coupling coefficient between oscillators 1 and 2 and given the close proximity of the resonators, it should be considered an effective coupling coefficient that takes into account interaction between the bright mode atom and its 2 nearest neighbors.

Of particular interest is the value of $\gamma_2$ which reflects the damping of the collective dark mode and plays a large role in dictating the linewidth of the EIT-like resonance. The radiative damping term, $\gamma_{R2}$, is minimized due to the collective oscillations of the array, mediated by near-field coupling between the unit cells. One consequence of utilizing collective modes is that the value of $\gamma_{R2}$ is inversely proportional to the size of the array due to the fact that the discontinuity at the edges of the array allow for light leakage to free space[27,28]. While collective modes have been utilized in past demonstrations of plasmonic EIT metamaterials to realize extremely small $\gamma_{R2}$ values, the Q-factors have intrinsically been limited by the Ohmic loss in the metal. By replacing the resonator constituents with lossless dielectrics, the non-radiative damping term $\gamma_{NR2}$ can also be minimized, resulting in the large increase in the Q-factor and peak transmittance.

Equally important is the role of the coupling coefficient, $\kappa$, in realizing a high Q-factor resonance for this system. It has previously been shown that the slope of dispersion is inversely proportional to $\kappa^2$.[2,28] In the limit of $\gamma_2 \to 0$, reduction of $\kappa$ will result in a monotonic increase of the Q-factor until $\kappa = 0$ is reached, at which point the Fano resonance will vanish, leaving only the bright mode resonance. In our system, the magnetic field from the bright mode resonator is inducing the dark mode resonance, as illustrated in Fig. 2a. Thus, as the spacing between the resonators ( $g_1$ ) is increased the coupling coefficient decreases, resulting in an increase in the Q-factor (Fig. 2b) and strong magnetic and electric field localization (Fig. 2c). At a spacing of $g_1 = 74$ nm, the Q-factor reaches a value of ~30,000 which is indicative of the fact that both radiative and non-radiative losses in the system are minimal, resulting in a situation wherein $\gamma_2 \sim 0$. However, losses in the Si primarily arising from surface states created during etching will ultimately limit the Q-factor, as has been demonstrated in microcavities[35]. Such losses are not included in these calculations. It is also important to realize that the bright mode resonator is also interacting with the dark mode resonator in the adjacent unit cell (D2), inducing a magnetic field that is 180° out of phase with the excitation from the bright mode resonator within its own unit cell. Therefore, when the gap between the bright mode resonator and adjacent dark mode resonators are equal ($g_1 = 75$ nm, $\Delta = g_2 - g_1 = 0$) the fields destructively interfere and $\kappa$ goes to zero, resulting in elimination of the Fano resonance, as can be observed in Fig. 2b and in the magnetic and electric field profiles in Fig. 2c.

At the transmission peak the energy in the array is concentrated in the collective dark mode with the magnetic dipoles in each of the split-rings oscillating coherently as shown in Fig. 3a. The array size thus becomes an important factor in the overall Q-factor of the metasurface due to the fact that lattice perturbations at the array's edge break the coherence[27,28], leading to strong scattering of light into free-space and broadening of the resonance peak. To characterize this effect, we fabricated arrays of five different sizes, consisting of 400 to 90,000 unit cells (SEM image of the test samples are shown in Fig. 3b). In the measurement, we used a 5x objective for illumination and a 50x objective for collection with an aperture in the conjugate image plane on the transmission side to confine the collection area to 15 μm x 15 μm. The measured spectra are shown in Fig. 3c and we observe that the Q-factor increases with increasing array size, saturating as we approach 90,000 unit cells. The saturation of the Q-factor indicates the limit imposed by both incoherent radiative loss arising from inhomogeneities and non-radiative loss arising due to finite absorption in the Si. As expected, the required unit cell number needed to achieve spectral convergence is larger than that reported in coherent plasmonic metamaterials due to the reduction in the non-radiative component.

Photonic crystal cavities possessing higher Q-factors have been reported[36], though they lack the spatial homogeneity present in the metasurface outlined here. Diffractive guided mode structures can also exhibit extremely high Q-factors[37], though in our case we have the additional freedom of engineering the local field enhancement through modification of the dark mode resonators. For instance, this can be done by placing symmetric feed-gaps in the resonator forming a split-ring, as shown in Fig. 4a. In this case, the normally oriented field located in the gap is enhanced by a factor of $\varepsilon_d / \varepsilon_s$ (Fig. 4b), where $\varepsilon_d$ and $\varepsilon_s$ are the permittivities of the resonator and surrounding medium, respectively. For the design depicted in Fig. 4, this results in an electric field enhancement of 44 in the gap region. Having the advantages of both a sharp spectral response and a strongly enhanced field in the surrounding substance allows the metasurface to serve as an ideal platform for enhancing interaction with the surrounding medium. To investigate the response of such structures, samples were made using the same processes as described above and measured in $n = 1.44$ index matching oil. An SEM image of the sample is shown in Fig. 4c. The simulated and experimentally measured transmittance curves are shown in Fig. 4d and result in theoretical and experimentally measured Q-factors of 374 and 129, respectively. The lower theoretical Q-factor, compared with the ring geometry, is due to the fact that the electric fields are not equal the two arms of the split-ring due to the asymmetry of the unit cell. This results in radiation loss to free space as the electric dipole moments do not fully cancel one another. The decrease in the experimental Q-factor, compared to theory, is attributed to more imperfections in the sample arising from the inclusion of the gaps, ultimately resulting in increased scattering from the structure.

**Refractive Index Sensing.** Due to their narrow linewidths, one interesting application of these metasurfaces is optical sensing. Optically resonant sensors are characterized by both the linewidth of the resonance ($\Delta\lambda$) as well as the shift in the resonance per refractive index unit change ($S$). These two values comprise the *FOM* which is given by $FOM = S/\Delta\lambda$ [38]. The highest demonstrated *FOMs* for Fano-resonant localized surface plasmon resonance (LSPR) sensors are on the order of 20[13,38,39]. Here, we examine the *FOM* of the ring resonator metasurfaces and to further increase the sensitivity of our metasurface to local index changes, we etched a 100-nm-tall quartz post below the Si resonators such that the field overlap in the surrounding dielectric can be further promoted, as illustrated in Fig. 5a. The transmittance spectra of the metasurface when immersed in oil with different refractive indices, from 1.40 to 1.44, are presented in Fig. 5b, and it can be seen that a substantial movement in the peak position is realized despite the small index change. A linear fit of the shift in the resonance peak (Fig. 5c) leads to a sensitivity of $S$ =289 nm/RIU, which is comparable, but slightly lower, than the best Fano-resonant plasmonic sensors. The slight decrease in sensitivity is due to the fact that the index contrast, and corresponding field enhancement, at the dielectric metasurface-oil interface is still lower than that found in plasmonic structures. However, when combining the sensitivity with an average resonance peak linewidth $\Delta\lambda$ of only 2.8 nm, we arrive at an *FOM* of 103, which far exceeds the current record for Fano-resonant LSPR sensors. With further optimization of fabrication and more stringent control of the resonator coupling, *FOMs* on the order of 1000 should be within reach. We also characterized the performance of the double-gap split-ring metasurface, finding it has a higher sensitivity ($S$ = 370 nm/RIU) due to the increased modal overlap with the surrounding material. However, the experimentally measured FOM is reduced to 37 due to the broader linewidth. More details can be found in the Supplementary Information section 3.

In conclusion, dielectric metasurfaces can be used to significantly improve upon the performance of their plasmonic counterparts in mimicking electromagnetically induced transparency due to their greatly reduced absorption loss, resulting in sensing *FOMs* that far exceed previously demonstrated LSPR sensors. With proper design, such metasurfaces can also confine the optical field to nanoscale regions, opening the possibility of using such metasurfaces for a wide range of applications including bio/chemical sensing, enhancing emission rates, optical modulation, and low-loss slow light devices.

**Methods**

**Sample fabrication.** Poly-crystalline Si was chosen as the resonator material due to its high index and small absorption in the near infrared frequencies. The Si was deposited on a 4 inch quartz wafer using a low pressure chemical vapor deposition (LPCVD) horizontal tube furnace. The metasurfaces were created by a sequential process of electron beam lithography (EBL), mask deposition, lift-off and reactive-ion etching (RIE). In the EBL process, we employed a 10 nm chromium charge dissipation layer on top of the PMMA resist. The total pattern size was ~225 μm x 225 μm, written using a JEOL 9300FS 100kV EBL tool. A fluorine-based ICP RIE recipe involving C4F8, SF6, O2 and Ar gas flows was used to etch poly-Si layer creating the metasurface layer. Polydimethylsiloxane (PDMS) flow cells were used to immerse the metasurface in the refractive index matching oils (Cargille Labs) for all the spectroscopy and sensing measurements. The samples were washed thoroughly after each measurement with Isopropyl Alcohol (IPA) followed by placing them in a vacuumed desiccator for 1 hour to ensure that no residual oil was left after each measurement.

**Optical measurement set-up.** A custom-built free-space infrared microscope was used to measure the transmittance, reflectance, and absorption spectra of the metasurface. The white light coming out from a Tungsten/Halogen lamp was first illuminated on the sample through a long working distance objective (Mitutoyo, 5x, 0.14 NA) with the numerical aperture (NA) of the objective further cut down to 0.025 by an aperture at its back focal plane, in order to restrict the incident angle to be close to normal. The transmittance spectrum (T) was normalized with the transmittance of the white light through an un-patterned quartz substrate. The reflectance spectrum (R) was normalized to reflectance of a gold mirror, and the absorption of the sample is simply determined as $A = 1 - T - R$.

**Acknowledgements**

This work was funded by the Office of Naval Research (ONR) under program N00014-12-1-0571. A portion of this research was conducted at the Center for Nanophase Materials Sciences, which is sponsored at Oak Ridge National Laboratory by the Scientific User Facilities Division, Office of Basic Energy Sciences, U.S. Department of Energy.

**Competing Financial Interests**

The authors declare no competing financial interests.

**Figures and Legends**

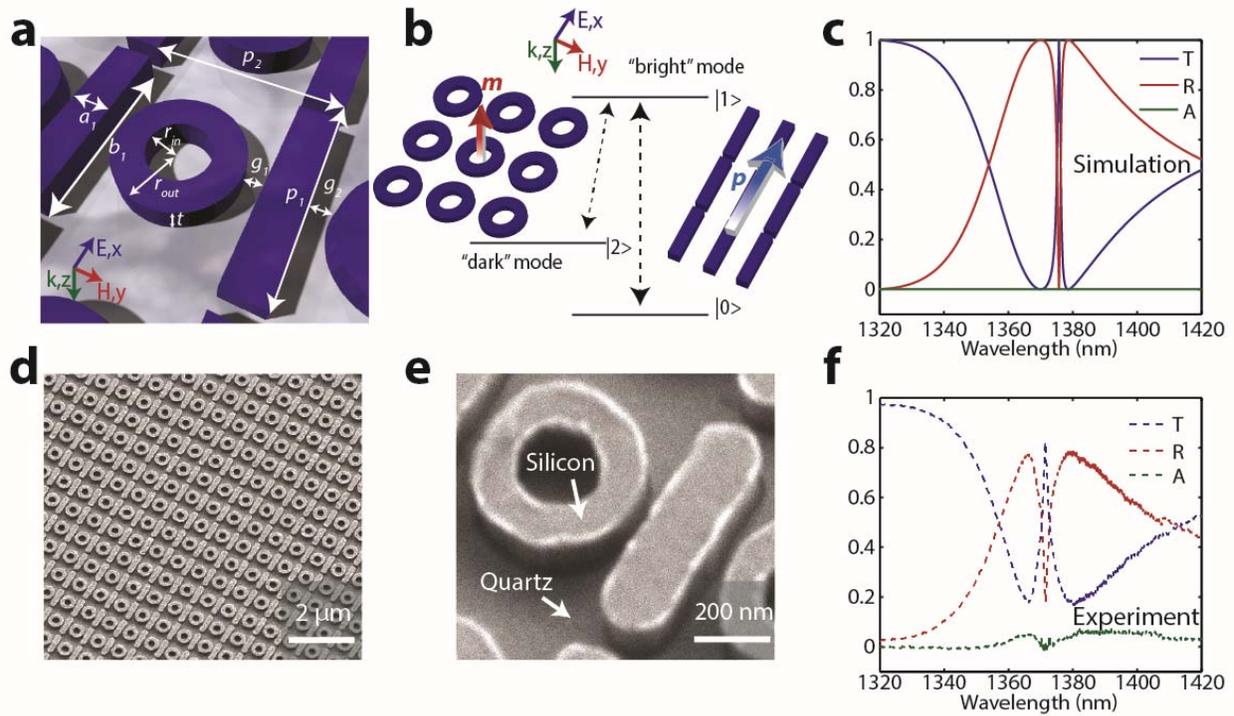

**Figure 1 Configuration and performance of the EIT metasurface.** a, Diagram of the metasurface. The geometrical parameters are: $a_1$ = 150 nm, $b_1$ = 720 nm, $r_{in}$ = 110 nm, $r_{out}$ = 225 nm, $g_1$ = 70 nm, $g_2$ = 80 nm, $t$ = 110nm, $p_1$ = 750 nm and $p_2$ = 750 nm. b, Schematic of interference between the bright and dark mode resonators. c, Simulated transmittance (blue curve), reflectance (red curve), and absorption (green curve) spectra of the metasurface. d, Oblique scanning electron microscope image of the fabricated metasurface. e, Enlarged image of a single unit cell. f, Experimentally measured transmittance, reflectance, and absorption spectra of the metasurface.

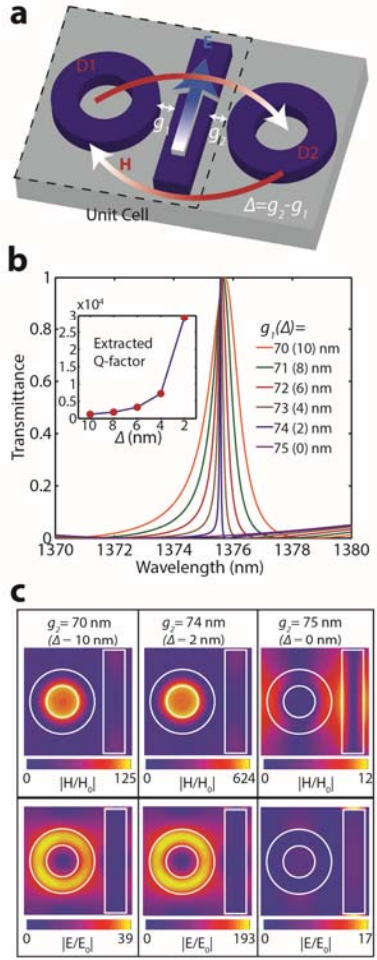

**Figure 2 Dependence of the Q-factor on the bright and dark mode resonator spacing.** a, Schematic showing the coupling of the bright mode resonator to the neighboring dark mode resonator. Here, we schematically provide the fields arising due to the bright mode resonance. b, Transmittance curve for the metasurface with varying values of $g_1$ obtained through FEFD simulation. $\Delta = g_2 - g_1$ is used to track the difference in spacing between the bright mode resonator and the adjacent dark mode resonators. The inset provides the extracted Q-factors of the EIT resonance as a function of $\Delta$. c, Simulated magnetic and electric field amplitudes as a function of resonator spacing.

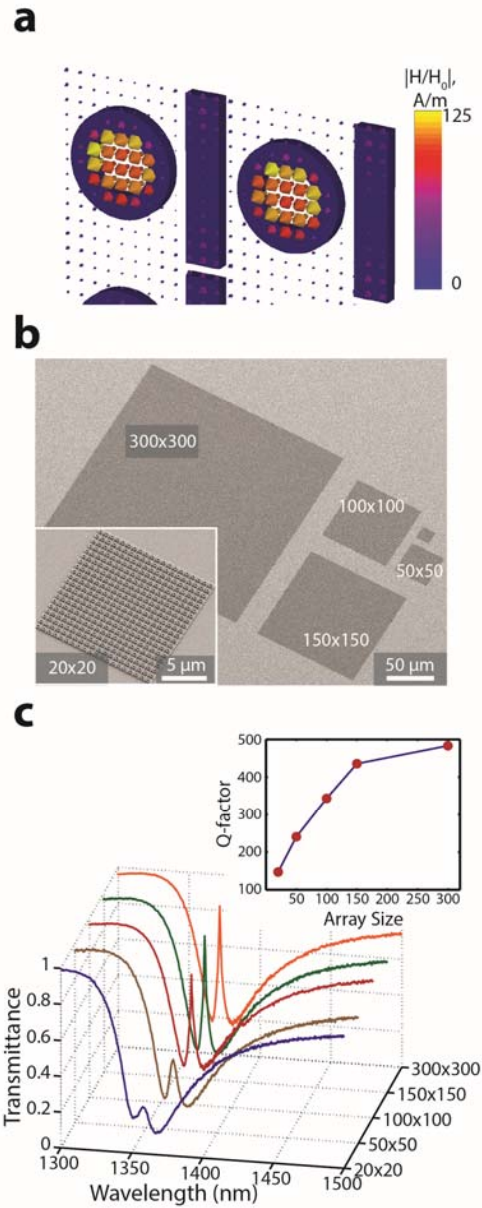

**Figure 3 Dependence of the Q-factor on the array size of the metasurface.** a, Vector plot of the magnetic field showing the coherent excitation of the magnetic dipoles within the dark mode resonators. b, Scanning electron microscope images of the dielectric metasurface with different array sizes. The inset shows the magnified view of a 20 by 20 array. c, Transmittance spectra of arrays with different sizes. The inset provides the extracted Q-factors of the metasurface as a function of array size.

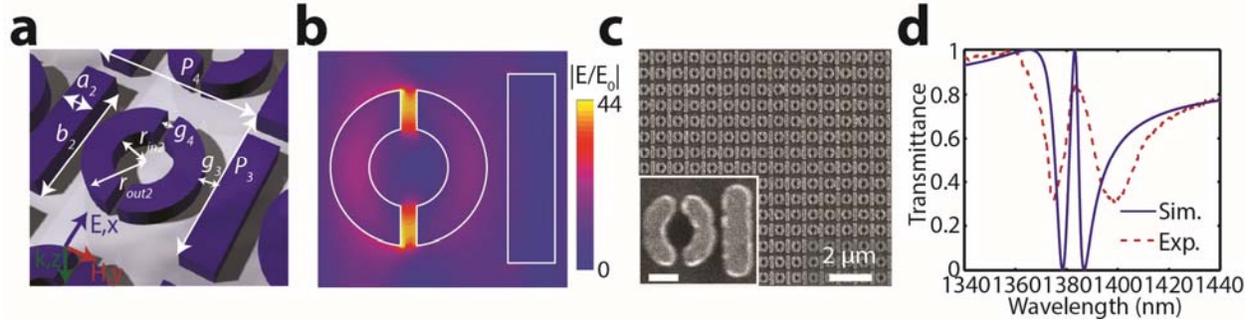

**Figure 4 Double-gap split-ring for enhancing E-field in the surrounding environment.** a, Diagram of the metasurface. The geometrical parameters are: $a_2$ = 150 nm, $b_2$ = 600 nm, $r_{in2}$ = 120 nm, $r_{out2}$ = 250 nm, $g_3$ = 70 nm, $g_4$ = 50 nm, $t$ = 110nm, $p_3$ = 750 nm and $p_4$ = 800 nm. b, Simulated electric field amplitude distribution. c, Scanning electron microscope image of the fabricated metasurface. The inset shows a close view of one single unit cell (scale bar of 200 nm). d, Simulated (blue solid line) and measured (red dashed line) transmittance spectrum of the metasurface.

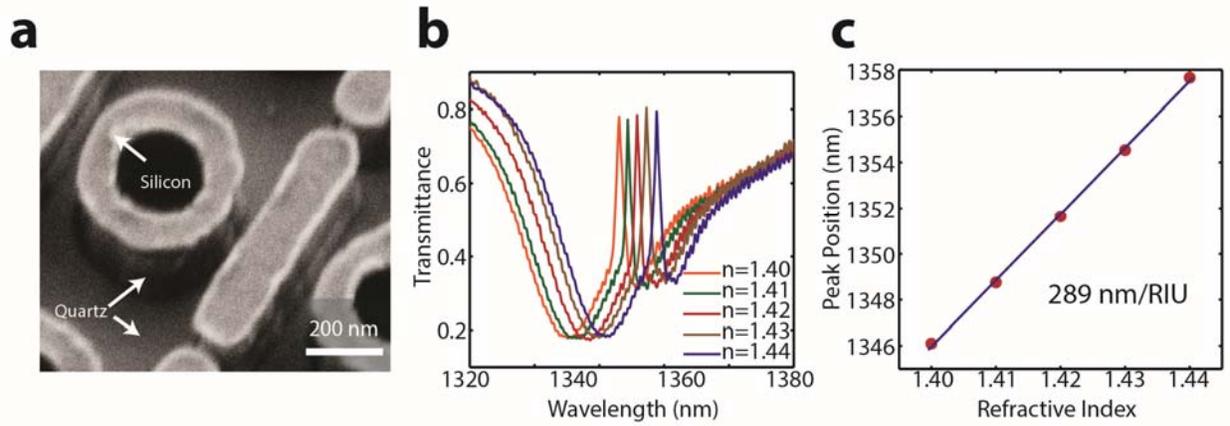

**Figure 5 Refractive Index Sensing.** a, SEM image of the sample with the quartz substrate etched by 100 nm. b, Measured transmittance spectra of the EIT metasurface when immersed in oil with refractive index ranging from 1.40 to 1.44. c, The red circles show the experimentally measured resonance peak position as a function of the background refractive index. The blue curve is a linear fit to the measured data, which was used to determine the sensitivity as 289 nm/RIU.